\documentclass[journal, column]{IEEEtran}\IEEEoverridecommandlockouts

\usepackage{amssymb}
\usepackage{amsmath}
\usepackage{amsfonts}
\usepackage{graphicx}
\usepackage{algorithm}
\usepackage{algorithmic}
\usepackage{epstopdf}
\usepackage{cite}
\usepackage{stfloats}
\usepackage{color}
\usepackage{subfigure}

\newtheorem{theorem}{\textbf{Theorem}}

\newtheorem{lemma}{\textbf{Lemma}}

\makeatletter
\renewcommand\normalsize{%
   \@setfontsize\normalsize\@xpt\@xiipt
   \abovedisplayskip 0.025\p@ \@plus0.05\p@ \@minus0.125\p@
   \abovedisplayshortskip \z@ \@plus0.075\p@
   \belowdisplayshortskip 0.15\p@ \@plus0.075\p@ \@minus0.025\p@
   \belowdisplayskip \abovedisplayskip
   \let\@listi\@listI}
\makeatother
\begin{document}

\title{\LARGE Low Complexity Algorithms for Robust Multigroup Multicast Beamforming}
\author{Guangda~Zang,\thanks{
%
%
Manuscript received January 19, 2019; revised April 19, 2019; accepted May 8, 2019.
This work was supported in part by the National Natural Science Foundation of China under Grant 61401272, Grant 61771309, Grant 61671301, Grant 61420106008, and Grant 61521062, in part by the Shanghai Key Laboratory Funding under Grant STCSM15DZ2270400.
The associate editor coordinating the review of this paper and approving it for publication was George C. Alexandropoulos. \emph{(Corresponding author: Feng Yang, e-mail: yangfeng@sjtu.edu.cn.)}

G. Zang, Y. Cui, W. Liu, F. Yang, L. Ding, and H. Liu are with the Department of Electronic Engineering, Shanghai Jiao Tong University, Shanghai 200240, China.

H. V. Cheng is with The Edward S. Rogers Sr. Department of Electrical and Computer Engineering, University of Toronto, Toronto  ON M5S 3G4, Canada.
}~Hei~Victor~Cheng,~Ying~Cui,~Wei~Liu,~Feng~Yang,~Lianghui~Ding,~and~Hui~Liu}
\maketitle

\begin{abstract}
Existing methods for robust multigroup multicast beamforming obtain feasible points using semidefinite relaxation (SDR) and Gaussian randomization, and have high computational complexity.
In this letter, we consider the robust multigroup multicast beamforming design to minimize the sum power (SP) or per-antenna power (PAP) under the signal-to-interference-plus-noise ratio (SINR) constraints and to maximize the worst-case SINR under the SP constraint or PAP constraints, respectively.
The resulting optimization problems are challenging non-convex problems with infinitely many constraints. For each problem, using the majorization-minimization (MM) approach, we propose an iterative algorithm to obtain a feasible solution which is shown to be a stationary point under certain conditions.
We also show that the proposed algorithms have much lower computational complexity than existing SDR-based algorithms.
\end{abstract}


\vspace{-0.2cm}
\section{Introduction}
\IEEEPARstart{W}{ith} multiple antennas equipped at the base station (BS), multicasting to different users in the form of beamforming is a promising way of increasing the throughput of a multicast system.
The single group multicast beamforming problem was first proposed and shown to be NP-hard in~\cite{multicast2006}. A semidefinite relaxation (SDR)-based method together with Gaussian randomization was proposed to obtain a feasible solution.
Later in~\cite{karipidis2008} the technique was extended to the multigroup case. SDR-based methods suffer from high computational complexity, prohibiting their applications in real systems. To reduce the computational complexity, majorization-minimization (MM)-based methods were then proposed in~\cite{SCA2014} and~\cite{tao2016}.

A common limitation of the aforementioned works is the perfect channel state information (CSI) assumption. In practice, channel estimates are noisy versions of the true channels. Thus, designing multicast beamforming under channel mismatch is of practical importance.
Worst-case robust multigroup multiuser beamforming was considered in~\cite{chen2012} under the sum power (SP) constraint and in~\cite{chris2014} under the per-antenna power (PAP) constraints.
However, both~\cite{chen2012} and~\cite{chris2014} apply  SDR-based methods to obtain feasible solutions, with high computational complexity.

To the best of our knowledge, low complexity solutions for the worst-case robust multigroup multicast beamforming problems are still unknown, and this work aims to fill this gap.
Our contribution lies in extending the MM-based methods for the perfect CSI case in~\cite{SCA2014,tao2016} to the case of channel uncertainty.
Specifically, we consider the robust multigroup multicast beamforming design to minimize the SP or PAP under the signal-to-interference-plus-noise ratio (SINR) constraints and to maximize the worst-case SINR under the SP constraint or PAP constraints, respectively. For each problem, using the MM approach~\cite{7547360}, we develop an algorithm to obtain a feasible point which is shown to be a stationary point under certain conditions. We also show that the proposed algorithms have much lower computational complexity than the existing SDR-based methods~\cite{chen2012,chris2014}. Numerical results demonstrate the advantages of the proposed algorithms over existing methods.

\section{System Model}
We consider a MIMO multicast system where a BS with $M~(\ge\! 1)$ antennas is serving $G~(\ge\! 1)$ multicast groups of single-antenna users.
Assume that every user belongs to only one (multicast) group.
Let $\mathcal{N}_g$ denote the set of indices of the users in group $g\!\in\!\mathcal{G}\!\triangleq\!\{1,\ldots,G\}$.
Let $N_g\triangleq |\mathcal{N}_g|$ denote the number of users in group $g\in\mathcal{G}$ and let $N_u\triangleq\sum\limits\nolimits_{g\in\mathcal{G}}N_g$ denote the total number of users.
Denote by ${{s}_{g}}\in \mathbb{C}$ the unit power data symbol intended for group $g$, and denote by ${{\mathbf{w}}_{g}}\in {\mathbb{C}^{M\times 1}}$ the beamforming vector applied to ${{s}_{g}}$. Let $\mathbf{w}\triangleq(\mathbf{w}_g)_{g\in\mathcal{G}}$.
The received signal of user $i\in\mathcal{N}_g,g\in\mathcal{G}$ is:
\begin{equation}
{y}_{ig}=\mathbf{h}_{ig}^{H}\mathbf{w}_{g}{s}_{g}+\sum\limits_{l\in\mathcal{G},l\ne g}{\mathbf{h}_{ig}^{H}{{\mathbf{w}}_{l}}{{s}_{l}}}+{{n}_{ig}},
\end{equation}
where ${{n}_{ig}}\!\sim\!\mathcal{CN}(0,\sigma _{ig}^{2})$ represents the independent additive white Gaussian noise (AWGN), $\mathbf{h}_{ig}^{H}\!\in\! \mathbb{C}^{1\times M}$ denotes the complex channel vector between the BS and user $i\!\in\!\mathcal{N}_g,g\!\in\!\mathcal{G}$, and $[\ \cdot\ ]^H$ denotes the Hermitian transpose of the argument.
In practice, only a noisy estimate of the true channel, denoted as ${{\mathbf{\hat{h}}}_{{ig}}}$, is available at the BS, and the mismatch can be modeled as ${{\mathbf{h}}_{ig}}\!=\!{{\mathbf{\hat{h}}}_{ig}}+{{\mathbf{e}}_{ig}}$, where $\mathbf{e}_{ig}\!\in\!\mathbb{C}^{M\times 1}$ denotes the corresponding error. Denote $\mathbf{e}\triangleq(\mathbf{e}_{ig})_{i\in\mathcal{N}_g,g\in\mathcal{G}}$.
Here we consider the widely used elliptic bounded error model, i.e., $\mathbf{e}_{ig}^{H}{{\mathbf{C}}_{ig}}{{\mathbf{e}}_{ig}}\le 1$, where ${{\mathbf{C}}_{ig}}\in\mathbb{S}^{M}$ is a Hermitian positive definite matrix which specifies the size and shape of the ellipsoid bound~\cite{chen2012,chris2014}.
Denote $\mathcal{S}_{ig}\triangleq\{\mathbf{e}_{ig}~|~ \mathbf{e}_{ig}^{H}{{\mathbf{C}}_{ig}}{{\mathbf{e}}_{ig}}\le 1\}$.
Assuming that each user decodes its requested data by treating interference as noise, the SINR at user~$i\in\mathcal{N}_g,g\in\mathcal{G}$ is:
\begin{equation}
\mathrm{SINR}_{ig}(\mathbf{w},\mathbf{e})=\frac{\big|\mathbf{w}_{g}^{H}(\mathbf{\hat{h}}_{ig}+\mathbf{e}_{ig})\big|^2}
{\sum\limits_{l\in\mathcal{G},l\ne g}{{{\big| \mathbf{w}_{l}^{H}({{{\mathbf{\hat{h}}}}_{ig}}+{{\mathbf{e}}_{ig}})\big|}^{2}}+\sigma _{ig}^{2}}}.\label{eq_sinr}
\end{equation}
Two commonly used metrics for power consumption of beamforming schemes are the SP radiated by the entire antenna array, ${P}_{s}(\mathbf{w})\triangleq\sum\limits\nolimits_{g\in\mathcal{G}}{\mathbf{w}_{g}^{H}{{\mathbf{w}}_{g}}}$,
and the PAP radiated by each antenna, ${{P}_{m}}(\mathbf{w})={{\big[ \sum\limits\nolimits_{g\in\mathcal{G}}{{{\mathbf{w}}_{g}}\mathbf{w}_{g}^{H}} \big]}_{mm}},~m=1,\ldots,M$,
where $\left[\ \cdot\ \right]_{mm}$ denotes the $m$-th diagonal element of the argument.
In the following, we shall study robust multigroup multicast beamforming design under imperfect CSI at the BS, considering the two power consumption metrics. It will be seen that the proposed solution framework can handle both metrics.

\section{Power Minimization With SINR Constraints}\label{sec_qos}
In this section, we consider the robust multigroup multicast beamforming design to minimize the SP or PAP under the SINR constraints. Specifically, we have:
\begin{align}
 \mathcal {P}_{\text{PM}}\!:\!\min_{r,\mathbf{w}}~&r \nonumber\\
  \mathrm{s.t.}~&\mathrm{SINR}_{ig}(\mathbf{w},\mathbf{e})\ge{{\tau }_{g}},
 \mathbf{e}_{ig} \!\in \mathcal{S}_{ig},i\!\in\mathcal{N}_g,g\in\mathcal{G}, \label{SINR_constraint} \\
 &{P}_{s}(\mathbf{w})\le r~(\text{or}~{{P}_{m}}(\mathbf{w})\le r, m=1,\ldots,M). \label{eq_power_cite}
\end{align}
Here, ${{\tau }_{g}}$ represents the minimum SINR requirement of group~$g$. Note that under perfect CSI, i.e., $\mathcal{S}_{i,g}\!=\!\{\mathbf{0}\},i\!\in\!\mathcal{N}_g,g\!\in\!\mathcal{G}$, Problem~$\mathcal {P}_{\text{PM}}$ has been shown to be NP-hard~\cite{multicast2006}. Under imperfect CSI, Problem~$\mathcal {P}_{\text{PM}}$ is even more challenging, as there are infinitely many constraints in~\eqref{SINR_constraint}. Problem~$\mathcal {P}_{\text{PM}}$ can be infeasible under certain $\tau_g,g\in\mathcal{G}$.
In this letter, we focus on the case that Problem~$\mathcal {P}_{\text{PM}}$ is feasible.

Existing methods for Problem~$\mathcal {P}_{\text{PM}}$ obtain feasible points using SDR and Gaussian randomization, and have high computational complexity~\cite{chen2012,chris2014}.
Specifically, the computational complexities of each iteration of an interior-point method used for solving one relaxed semidefinite programming for the SP minimization~\cite{chen2012} and for the PAP minimization~\cite{chris2014} are $\mathcal{O}(\max\{G^3M^6,G^2M^4N_u\})$ and $\mathcal{O}(\max\{G^3M^6,G^2M^4(N_u\!+\!M)\})$, respectively.
The worst-case computational complexity of each Gaussian randomization is $\mathcal{O}(G^{3.5})$, and usually 100 Gaussian randomizations have to be conducted to obtain a feasible solution with promising performance.
In the following, we develop an algorithm of much lower computational complexity to obtain feasible solutions of Problem~$\mathcal {P}_{\text{PM}}$ with desirable performance, which can be shown to be stationary points under certain conditions.

Specifically, to tackle the challenge caused by~\eqref{SINR_constraint}, we first have the following result.
\begin{lemma}\label{lemma_SINR}
For all $G\ge 1$ and all $\mathbf{C}_{ig}\in\mathbb{S}^{M},i\in\mathcal{N}_g,g\in\mathcal{G}$, if $\mathbf{w}$ satisfies
\begin{align}
&\zeta(\mathbf{w},\tau_g)-\big| \mathbf{w}_{g}^{H}{{{\mathbf{\hat{h}}}}_{ig}} \big| \le 0,~i\in\mathcal{N}_g,g\in\mathcal{G},&\label{SINR_constraint_convex}
\end{align}
then it satisfies~\eqref{SINR_constraint},
where $\zeta(\mathbf{w},\tau_g)\triangleq{\varepsilon_{ig}}\big\| {{\mathbf{w}}_{g}} \big\|_2+\sqrt{{\tau }_{g}}\sqrt{\sum\limits_{l\in\mathcal{G},l\ne g}{{{\big( \big| \mathbf{w}_{l}^{H}{{{\mathbf{\hat{h}}}}_{ig}} \big|+{\varepsilon_{ig}}\big\| {{\mathbf{w}}_{l}} \big\|_2 \big)}^{2}}+\sigma _{ig}^{2}}}$.
Furthermore, for $G=1$ and ${{\mathbf{C}}_{i1}}=1/\mu^{2}{{\mathbf{I}}_{M}}$,\footnote{Note that ${{\mathbf{C}}_{i1}}=1/\mu^{2}{{\mathbf{I}}_{M}}$ corresponds to the sphere bounded error model with radius ${\mu}$, which is also widely used~\cite{voro2003}.} where $\mu$ is a positive constant and $\mathbf{I}_M$ denotes the $M\times M$ identity matrix,
if $\mathbf{w}$ satisfies~\eqref{SINR_constraint}, then it satisfies~\eqref{SINR_constraint_convex}.
\end{lemma}
\begin{IEEEproof}
Using triangle inequality, we have:
\begin{align}
\big| {{\mathbf{w}}_g^{H}}\mathbf{\hat{h}}_{ig}+{{\mathbf{w}}_g^{H}}\mathbf{e}_{ig} \big|\ge \big| {{\mathbf{w}}_g^{H}}\mathbf{\hat{h}}_{ig} \big|-\big| {{\mathbf{w}}_g^{H}}\mathbf{e}_{ig} \big|,\label{eq_cs_inequality1}\\
\big| {{\mathbf{w}}_g^{H}}\mathbf{\hat{h}}_{ig}+{{\mathbf{w}}_g^{H}}\mathbf{e}_{ig} \big|\le \big| {{\mathbf{w}}_g^{H}}\mathbf{\hat{h}}_{ig} \big|+\big| {{\mathbf{w}}_g^{H}}\mathbf{e}_{ig} \big|.\label{eq_cs_inequality1b}
\end{align}
By $\mathbf{e}_{ig}^{H}{{\mathbf{C}}_{ig}}{{\mathbf{e}}_{ig}}\le 1$ and Cauchy-Schwarz inequality, we have:
\begin{flalign}
 \big| {{\mathbf{w}}_g^{H}}\mathbf{e}_{ig}\big|{\le}
 \big\| \mathbf{w}_g \big\|_2\big\|\mathbf{e}_{ig}\big\|_2{\le}\!
 \frac{\big\| \mathbf{w}_g \big\|_2}
 {\sqrt{\operatorname{R}(\mathbf{C}_{ig},\mathbf{e}_{ig})}}
 {\le}\!
 \frac{\big\| \mathbf{w}_g \big\|_2}{\sqrt{\lambda_\text{min}(\mathbf{C}_{ig})}},\label{eq_rayleigh}
\end{flalign}
where $\operatorname{R}(\mathbf{C}_{ig},\mathbf{e}_{ig})\triangleq
{
\mathbf{e}_{ig}^{H}
\mathbf{C}_{ig}\mathbf{e}_{ig}
} \big/
{
\mathbf{e}_{ig}^H\mathbf{e}_{ig}
}$ is the Rayleigh quotient and
$\lambda_\text{min}(\mathbf{C}_{ig})$ denotes the smallest eigenvalue of $\mathbf{C}_{ig}$. Letting $\varepsilon_{ig}\triangleq1\big/\sqrt{{\lambda_\text{min}(\mathbf{C}_{ig})}}$ and by~\eqref{eq_cs_inequality1},~\eqref{eq_cs_inequality1b} and~\eqref{eq_rayleigh}, we have:
\begin{align}
\big| {{\mathbf{w}}_g^{H}}\mathbf{\hat{h}}_{ig}+{{\mathbf{w}}_g^{H}}\mathbf{e}_{ig} \big|\ge \big| {{\mathbf{w}}_g^{H}}\mathbf{\hat{h}}_{ig} \big|-{\varepsilon_{ig}}\big\| \mathbf{w}_g \big\|_2,\label{eq_cs_inequality2}\\
\big| {{\mathbf{w}}_g^{H}}\mathbf{\hat{h}}_{ig}+{{\mathbf{w}}_g^{H}}\mathbf{e}_{ig} \big|\le \big| {{\mathbf{w}}_g^{H}}\mathbf{\hat{h}}_{ig} |+{\varepsilon_{ig}}\big\| \mathbf{w}_g \big\|_2.\label{eq_cs_inequality3}
\end{align}
Taking square root on both sides of~\eqref{SINR_constraint}, and by~\eqref{eq_cs_inequality2} and~\eqref{eq_cs_inequality3},
we have~\eqref{SINR_constraint_convex}.
Furthermore, when $G=1$ and ${{\mathbf{C}}_{i1}}=1/\mu^{2}{{\mathbf{I}}_{M}}$, we have $\varepsilon_{i1}=\mu$ and the error bounds become $\left\| {{\mathbf{e}}_{{i1}}} \right\|_{2}^{2}\le \mu^{2},~i\in\mathcal{N}_1$.
It can be verified that $\big| {{\mathbf{w}}_1^{H}}\mathbf{\hat{h}}_{i1}+{{\mathbf{w}}_1^{H}}\mathbf{e}_{i1} \big|= \big| {{\mathbf{w}}_1^{H}}\mathbf{\hat{h}}_{i1} \big|-{\varepsilon_{i1}}\big\| \mathbf{w}_1 \big\|_2,~i\!\in\!\mathcal{N}_1$, if $\mathbf{e}_{i1}\!=\!-\varepsilon_{i1} e^{j\angle \mathbf{w}_{1}^H\mathbf{\hat{h}}_{i1}}\cdot\mathbf{w}_{1}\big/{\big\|\mathbf{w}_{1}\big\|_2}$~\cite{voro2003}. Thus,~\eqref{SINR_constraint_convex} is equivalent to~\eqref{SINR_constraint}.
\end{IEEEproof}

In addition,
we introduce slack variables $\mathbf{s}\triangleq(s_{ig})_{i\in\mathcal{N}_g,g\in\mathcal{G}}$ and add a penalty term in the objective to force $\mathbf{s}$ toward $0$. Then, we obtain an approximate problem of Problem~$\mathcal {P}_{\text{PM}}$:
\begin{flalign}
&\mathcal {P}_{\text{App-PM}}~:~\min_{r,\mathbf{w},\mathbf{s}}~r+\sum_{i\in\mathcal{N}_g,g\in\mathcal{G}}s_{ig} &\nonumber\\
&\quad\quad\quad\mathrm{s.t.}~\zeta(\mathbf{w},\tau_g)-\big| \mathbf{w}_{g}^{H}{{{\mathbf{\hat{h}}}}_{ig}} \big| \le s_{ig},i\in\mathcal{N}_g,g\in\mathcal{G},&\\
&\quad\quad\quad\quad~~s_{ig}\geq 0,i\in\mathcal{N}_g,g\in\mathcal{G},&\label{eq_s}\\
&\quad\quad\quad\quad~\ \eqref{eq_power_cite}.&\nonumber
\end{flalign}
Observe that Problem~$\mathcal {P}_{\text{App-PM}}$ is always feasible. If $(r,\mathbf{w},\mathbf{0})$ is a feasible solution of Problem~$\mathcal {P}_{\text{App-PM}}$, then $(r,\mathbf{w})$ is also a feasible solution of Problem~$\mathcal {P}_{\text{PM}}$.
Besides, the optimal value of Problem~$\mathcal {P}_{\text{App-PM}}$ is generally no smaller than that of Problem~$\mathcal {P}_{\text{PM}}$, and when $G=1$ and ${{\mathbf{C}}_{i1}}=1/\mu^{2}{{\mathbf{I}}_{M}}$, the optimal values of Problem~$\mathcal {P}_{\text{App-PM}}$ and Problem~$\mathcal {P}_{\text{PM}}$ are the same.
In the following, we focus on solving Problem~$\mathcal {P}_{\text{App-PM}}$ which has a simpler form than Problem~$\mathcal {P}_{\text{PM}}$.

Problem~$\mathcal {P}_{\text{App-PM}}$ is non-convex due to the non-convexity of~\eqref{SINR_constraint_convex}. We develop an algorithm (Algorithm~\ref{Alg_Opt}) to obtain a stationary point of Problem~$\mathcal {P}_{\text{App-PM}}$ using the MM approach in~\cite{7547360}. The main idea here is to successively solve a sequence of convex approximations of Problem~$\mathcal {P}_{\text{App-PM}}$.
Specifically, the approximate problem at iteration~$k$ is Problem~$\mathcal {P}_{\text{MM-PM1}}$ when $\mathbf{s}^{(k-1)}\neq \mathbf{0}$ and is Problem~$\mathcal {P}_{\text{MM-PM2}}$ when $\mathbf{s}^{(k-1)}= \mathbf{0}$:

\vspace{-0.2cm}
\begin{small}
\begin{flalign}
 &\mathcal {P}_{\text{MM-PM1}}:\big(\mathbf{w}^{(k)},r^{(k)},\mathbf{s}^{(k)}\big)\triangleq\arg\min_{r,\mathbf{w},\mathbf{s}}~r+\sum_{i\in\mathcal{N}_g,g\in\mathcal{G}}s_{ig} \nonumber&\\
 &\mathrm{s.t.}~\zeta(\mathbf{w},\tau_g){-}\frac{\operatorname{Re}\big( (\mathbf{w}_{g}^{(k-1)})^{H}{{{\mathbf{\hat{h}}}}_{ig}}\mathbf{\hat{h}}_{ig}^{H}{{\mathbf{w}}_{g}} \big)}{\big| (\mathbf{w}_{g}^{(k-1)})^{H}{{{\mathbf{\hat{h}}}}_{ig}} \big|}\le s_{ig},i{\in}\mathcal{N}_g,g{\in}\mathcal{G},\!\!\!\!\!\!& \label{eq_MM1_sinr}\\
 &~\quad~s_{ig}\geq 0,i\in\mathcal{N}_g,g\in\mathcal{G},&\\
 &~\quad~\eqref{eq_power_cite},\nonumber&
\end{flalign}
\end{small}
\begin{small}
\begin{flalign}
 &\mathcal {P}_{\text{MM-PM2}}:\big(\mathbf{w}^{(k)},r^{(k)},\mathbf{s}^{(k)}\big)\triangleq\arg\min_{r,\mathbf{w},\mathbf{s}}~r \nonumber&\\
 &\mathrm{s.t.}~\zeta(\mathbf{w},\tau_g){-}\frac{\operatorname{Re}\big( (\mathbf{w}_{g}^{(k-1)})^{H}{{{\mathbf{\hat{h}}}}_{ig}}\mathbf{\hat{h}}_{ig}^{H}{{\mathbf{w}}_{g}} \big)}{\big| (\mathbf{w}_{g}^{(k-1)})^{H}{{{\mathbf{\hat{h}}}}_{ig}} \big|}\le 0,i{\in}\mathcal{N}_g,g{\in}\mathcal{G},\!\!& \label{eq_MM1_sinr}\\
 &~\quad~s_{ig}=0,i\in\mathcal{N}_g,g\in\mathcal{G},&\\
 &~\quad~\eqref{eq_power_cite},\nonumber&
\end{flalign}\end{small}where $\mathbf{w}^{(k-1)}$ is an optimal solution of Problem~$\mathcal {P}_{\text{MM-PM1}}$ (Problem~$\mathcal {P}_{\text{MM-PM2}}$) at iteration~$k-1$.
\begin{lemma}\label{lemma_convex_app}
The constraints in~\eqref{eq_MM1_sinr} are convex.
\end{lemma}
\begin{IEEEproof}
For any given complex vector $\mathbf{u}$,
we have~\cite{7547360}
$\big| {{\mathbf{w}}_g^{H}}\mathbf{\hat{h}}_{ig} \big|\ge {\operatorname{Re}(\mathbf{u}^H\mathbf{\hat{h}}_{ig}{{\mathbf{\hat{h}}}^{H}}_{ig}\mathbf{w}_g)}\Big/{\big|\mathbf{u}^H\mathbf{\hat{h}}_{ig} \big|}.$
Define
$\mathbf{v}_{ig}{\triangleq}{{[q_{ig}( {{\mathbf{w}}_{1}} ),\ldots, q_{ig}( {{\mathbf{w}}_{g{-}1}} ), q_{ig}( {{\mathbf{w}}_{g{+}1}} ),\ldots,q_{ig}( {{\mathbf{w}}_{G}} ),{{\sigma }_{ig}}]}^{T}},$
where $q_{ig}( {{\mathbf{w}}_{l}}){=}\big( \big| \mathbf{w}_{l}^{H}{{{\mathbf{\hat{h}}}}_{ig}} \big|+{\varepsilon_{ig}}\big\| {{\mathbf{w}}_{l}} \big\|_2 \big)$. $\|\cdot\|_2$ is convex and non-decreasing in every argument, and $q_{ig}(\ \!\cdot\ \!)$ is a convex function. By
vector composition rule, $\|\mathbf{v}_{ig}\|_2$ is convex in $\mathbf{w}_g$. Since ${\varepsilon_{ig}}\left\| {{\mathbf{w}}_{g}} \right\|_2$ is convex in $\mathbf{w}_g$, ${\varepsilon_{ig}}\big\| {{\mathbf{w}}_{g}} \big\|_2\!+\!\sqrt{{\tau }_{g}}\big\|\mathbf{v}_{ig}\big\|_2$ is convex in $\mathbf{w}_g$.\end{IEEEproof}

Now, we analyze the convergence behavior of Algorithm~\ref{Alg_Opt}.
\begin{theorem}\label{theorem_qos}
For any initial point, Algorithm~\ref{Alg_Opt} converges to a stationary point of Problem~$\mathcal {P}_{\text{App-PM}}$,
which corresponds to a stationary point of Problem~$\mathcal {P}_{\text{PM}}$ when $G=1$, ${{\mathbf{C}}_{i1}}=1/\mu^{2}{{\mathbf{I}}_{M}}$ and $\mathbf{s}=\mathbf{0}$.
\end{theorem}
\begin{IEEEproof}
Let functions $\bar{f}_{ig}(\mathbf{w};\mathbf{w}^{(k-1)})$ and $f_{ig}(\mathbf{w})$ denote the left hand sides of~\eqref{eq_MM1_sinr} and~\eqref{SINR_constraint_convex}, respectively. It can be verified that $\bar{f}_{ig}(\mathbf{w}^{(k-1)};\mathbf{w}^{(k-1)})=f_{ig}(\mathbf{w}^{(k-1)})$ and $\bar{f}_{ig}(\mathbf{w};\mathbf{w}^{(k-1)})\ge f_{ig}(\mathbf{w})$, for all $\mathbf{w}$. Thus, $\bar{f}_{ig}(\mathbf{w};\mathbf{w}^{(k-1)})$ is a global upper bound of $f_{ig}(\mathbf{w})$, and touches it at $\mathbf{w}^{(k-1)}$.
By~\cite{7547360}, we can show that Algorithm~\ref{Alg_Opt} converges to a stationary point of Problem~$\mathcal {P}_{\text{App-PM}}$.
Moreover, when $G=1$, ${{\mathbf{C}}_{i1}}=1/\mu^{2}{{\mathbf{I}}_M}$ and $\mathbf{s}=\mathbf{0}$, by Lemma~\ref{lemma_SINR},
we can show that a stationary point of Problem~$\mathcal{P}_\text{App-PM}$
corresponds to a stationary point of Problem~$\mathcal{P}_\text{PM}$.\end{IEEEproof}

Next, we analyze the computational complexity of Algorithm~\ref{Alg_Opt}.
The computational complexities of each iteration of an interior-point method in Step~\ref{alg_qos_interior_points} used for solving Problem~$\mathcal{P}_\text{MM-PM}$ for the SP minimization and for the PAP minimization are $\mathcal{O}(\max\{ G^3M^3,G^2M^2N_u\})$ and $\mathcal{O}(\max\{ G^3M^3,G^2M^2(N_u+M)\})$, respectively.
Although the number of iterations of Algorithm~\ref{Alg_Opt} cannot be analytically characterized, numerical results show that Algorithm~\ref{Alg_Opt} terminates in a few iterations.
Therefore, we can conclude that Algorithm~\ref{Alg_Opt} has much lower computational complexity than the SDR-based methods in~\cite{chen2012},~\cite{chris2014},~especially for large~$M$.

\setlength{\textfloatsep}{0.1cm}
\begin{algorithm}[t]
\caption{Algorithm for Power Minimization Problem}
\label{Alg_Opt}
\begin{algorithmic}[1]
\small{
\renewcommand{\algorithmicrequire}{\textbf{Initialization:}}
\REQUIRE {Set $k=0$, and choose any ${s}^{(0)}_{ig}>0,i\in\mathcal{N}_g,g\in\mathcal{G}$ and $\mathbf{w}_g^{(0)}\in\mathbb{C}^{M\times 1},g\in\mathcal{G}$. Choose a threshold $\xi>0$.}
\REPEAT
\STATE Given $\mathbf{w}^{(k)}$, obtain $\big(\mathbf{w}^{(k+1)},{r}^{(k+1)},\mathbf{s}^{(k+1)} \big)$ by solving Problem~$\mathcal {P}_{\text{MM-PM1}}$ when $\mathbf{s}^{(k)}\neq \mathbf{0}$ or Problem~$\mathcal {P}_{\text{MM-PM2}}$ when $\mathbf{s}^{(k)}= \mathbf{0}$ using interior-point methods;\label{alg_qos_interior_points}
\STATE $k \gets k+1$;
\UNTIL $\big| {{r}^{(k)}}-{{r}^{(k-1)}} \big|\le \xi$.
}\normalsize
\end{algorithmic}
\end{algorithm}
\setlength{\floatsep}{0.1cm}

\section{Max-Min Fair Problem With Power Constraints}\label{sec_mmf}
In this section, we consider the robust multigroup multicast beamforming design to maximize the worst-case SINR under the SP constraint or PAP constraints. In particular, we have:
\begin{flalign}
&\mathcal {P}_{\text{MMF}}:\max_{t,\mathbf{w}}~\min_{i\in\mathcal{N}_g,g\in\mathcal{G}}\min_{\mathbf{e}_{ig}\in\mathcal{S}_{ig}}~\mathrm{SINR}_{ig}(\mathbf{w},\mathbf{e})& \nonumber\\
&\quad~~\mathrm{s.t.}~{P}_{s}(\mathbf{w})\le \gamma~(\text{or}~{{P}_{m}}(\mathbf{w})\le \gamma, m=1,\ldots,M),&\label{eq_power_cite_MMF}
\end{flalign}
where $\gamma$ represents the power limit.

Problem~$\mathcal {P}_{\text{MMF}}$ is always feasible and can be transformed to an epigraph form problem by introducing an auxiliary variable and infinitely many inequality constraints.
Existing methods for Problem~$\mathcal {P}_{\text{MMF}}$~\cite{chen2012,chris2014} obtain feasible solutions using bisection, with SDR and Gaussian randomization in each bisection step. The computational complexities of each bisection step for the SP constraint and PAP constraints are the same as those for the SP minimization in~\cite{chen2012} and PAP minimization in~\cite{chris2014}.
Therefore, the existing SDR-based methods~\cite{chen2012,chris2014} for Problem~$\mathcal {P}_{\text{MMF}}$ also have high computational complexity. In the following, we develop an algorithm of much lower computational complexity to obtain feasible solutions of Problem~$\mathcal {P}_{\text{MMF}}$ with desirable performance, which can be shown to be stationary points under certain conditions.

Specifically, by introducing an auxiliary variable $t$, imposing constraints $\mathrm{SINR}_{ig}(\mathbf{w},\mathbf{e})\ge t,\mathbf{e}_{ig} \!\in \mathcal{S}_{ig},i\!\in\!\mathcal{N}_g,g\!\in\!\mathcal{G}$, and then using Lemma~\ref{lemma_SINR}, we obtain an approximate problem of Problem~$\mathcal {P}_{\text{MMF}}$:
\begin{flalign}
&~\mathcal {P}_{\text{App-MMF}}:\max_{t,\mathbf{w}}~t  &\nonumber\\
&~\quad\quad\quad\quad\quad\mathrm{s.t.}~\zeta(\mathbf{w},t)-\big| {{\mathbf{w}}_g^{H}}\mathbf{\hat{h}}_{ig} \big| \le 0,i\in\mathcal{N}_g,g\in\mathcal{G},\!\!\!\!\!\!\!&\\
&~\quad\quad\quad\quad\quad\quad~~\eqref{eq_power_cite_MMF}.&\nonumber
\end{flalign}

Similarly, any feasible point of Problem~$\mathcal {P}_{\text{App-MMF}}$ is also a feasible point of Problem~$\mathcal {P}_{\text{MMF}}$.
Besides, the optimal value of Problem~$\mathcal {P}_{\text{App-MMF}}$ is in general no larger than that of Problem~$\mathcal {P}_{\text{MMF}}$, and when $G=1$ and ${{\mathbf{C}}_{i1}}=1/\mu^{2}{{\mathbf{I}}_{M}}$, the optimal values of Problem~$\mathcal {P}_{\text{App-MMF}}$ and Problem~$\mathcal {P}_{\text{MMF}}$ are same. Thus, we solve Problem~$\mathcal {P}_{\text{App-MMF}}$ which has a simpler form than Problem~$\mathcal {P}_{\text{MMF}}$.

Problem~$\mathcal {P}_{\text{App-MMF}}$ is non-convex.
We develop an algorithm (Algorithm~\ref{Alg_Opt_Max_Min_Fair_Problem}) of two loops to obtain a stationary point of Problem~$\mathcal {P}_{\text{App-MMF}}$. Specifically, in the outer loop, we adopt the MM approach~\cite{7547360}.
The approximation of Problem~$\mathcal {P}_{\text{App-MMF}}$ at iteration~$k$ is:

\begin{small}
\begin{flalign}
&\mathcal {P}_{\text{MM-MMF}}:\big(\mathbf{w}^{(k)},t^{(k)}\big)\triangleq\arg\max_{t,\mathbf{w}}~t \nonumber&\\
&\mathrm{s.t.}~\zeta(\mathbf{w},t)\!-\!\frac{\operatorname{Re}\big( (\mathbf{w}_{g}^{(k-1)})^{H}{{{\mathbf{\hat{h}}}}_{ig}}\mathbf{\hat{h}}_{ig}^{H}{{\mathbf{w}}_{g}} \big)}{\big| (\mathbf{w}_{g}^{(k-1)})^{H}{{{\mathbf{\hat{h}}}}_{ig}} \big|}\!\le \!0,i\in\mathcal{N}_g,g\in\mathcal{G},\!\!\!\!\!\!\!\!& \label{eq_MMF_SINRSCA}\\
&~\quad\ \eqref{eq_power_cite_MMF},&\nonumber
\end{flalign}
\end{small}where $\mathbf{w}^{(k-1)}$ is an optimal solution of Problem~$\mathcal {P}_{\text{MM-MMF}}$ at iteration~$k-1$.
The inequalities in~\eqref{eq_MMF_SINRSCA} are convex in $t$ and $\mathbf{w}_g$, respectively, but not in both simultaneously.
In fact, the sets defined by~\eqref{eq_MMF_SINRSCA} are quasi-convex.
Thus, different from Problem~$\mathcal {P}_{\text{MM-PM}}$ which is convex, Problem~$\mathcal {P}_{\text{MM-MMF}}$ is quasi-convex.
In the inner loop, we solve Problem~$\mathcal {P}_{\text{MM-MMF}}$ using bisection, where a convex feasibility problem is solved at each bisection step.
Note that $\mathbf{w}^{(0)}$ in Algorithm~\ref{Alg_Opt_Max_Min_Fair_Problem} is a feasible solution of Problem~$\mathcal{P}_\text{Feasibility}$.
Following the proof for Theorem~\ref{theorem_qos}, we now analyze the convergence behavior of Algorithm~\ref{Alg_Opt_Max_Min_Fair_Problem}.
\begin{theorem}
For any feasible initial point, Algorithm~\ref{Alg_Opt_Max_Min_Fair_Problem} converges to a stationary point of Problem~$\mathcal {P}_{\text{App-MMF}}$, which is also a stationary point of Problem~$\mathcal {P}_{\text{MMF}}$ when $G=1$ and ${{\mathbf{C}}_{i1}}=1/\mu^{2}{{\mathbf{I}}_{M}}$.
\end{theorem}

Next, we analyze the computational complexity of Algorithm~\ref{Alg_Opt_Max_Min_Fair_Problem}.
Numerical results show that the outer loop of Algorithm~\ref{Alg_Opt_Max_Min_Fair_Problem} terminates in a few iterations.
There are $\left \lceil \log_2\big((t_U\!-\!t_L)/\xi_1\big)  \right \rceil$ bisection steps in the inner loop, where $\left \lceil \ \cdot\  \right \rceil$ denotes the ceiling function.
In addition, in each bisection step,
the computational complexities of each iteration of an interior-point method used for solving Problem~$\mathcal{P}_\text{Feasibility}$ under the SP constraint and under the PAP constraints are $\mathcal{O}(\max\{ G^3M^3,G^2M^2N_u\})$ and $\mathcal{O}(\max\{ G^3M^3,G^2M^2(N_u+M)\})$, respectively.
Therefore, we know that the computational complexity of Algorithm~\ref{Alg_Opt_Max_Min_Fair_Problem} is much lower than those of the SDR-based methods in~\cite{chen2012,chris2014}.

\begin{algorithm}[t]
\caption{Algorithm for Max-Min Fair Problem}
\label{Alg_Opt_Max_Min_Fair_Problem}
\begin{algorithmic}[1]
\small{
\renewcommand{\algorithmicrequire}{\textbf{Initialization:}}
\REQUIRE {Set $k=0$ and choose any $\tilde{\mathbf{w}}_g\in\mathbb{C}^{M\times 1}, g\in\mathcal{G}$. Set $\mathbf{w}_g^{(0)}=\frac{\sqrt{\gamma}}{\sqrt{G}}\frac{\tilde{\mathbf{w}}_g}{\|\tilde{\mathbf{w}}_g\|_2}$ for the SP constraint or $\mathbf{w}_g^{(0)}=\frac{\sqrt{\gamma}}{\sqrt{G}}\frac{\tilde{\mathbf{w}}_g}{\|\tilde{\mathbf{w}}_g\|_\infty}$ for the PAP constraints, $g \in\mathcal{G}$. 
Choose thresholds $\xi_1>0$ and $\xi_2>0$. Choose $t_L\ge0$ and $t_U>t_L$.}
\REPEAT
\REPEAT
\STATE{$t:={(t_L+t_U)}\big/{2}$ and for given $\mathbf{w}^{(k)}$, solve
\begin{equation}
\mathcal{P}_\text{Feasibility}:\text{find}~\mathbf{w}\quad\quad
\mathrm{s.t.}~\eqref{eq_MMF_SINRSCA},~\eqref{eq_power_cite_MMF},\nonumber
\end{equation}
\\using interior-point methods;}\label{algo_MMF_feasibility}
\STATE {\textbf{if} Problem~$\mathcal {P}_{\text{Feasibility}}$ is feasible, $t_L := t$; \textbf{else} $t_U := t$;}
\UNTIL {$t_U-t_L\le \xi_1 $.}
\STATE {Set $\mathbf{w}^{(k+1)}=\mathbf{w}$ and ${{t}^{(k+1)}} =t$;}
\STATE {$k \gets k+1$;}
\UNTIL $\big| {{t}^{(k)}}-{{t}^{(k-1)}} \big|\le \xi_2$.
}\normalsize
\end{algorithmic}
\end{algorithm}

\section{Numerical Results}\label{Sec_simulation}
In this section, we show the performance of Algorithm~\ref{Alg_Opt_Max_Min_Fair_Problem} for Problem~$\mathcal{P}_\text{MMF}$ under the SP constraint and the PAP constraints, respectively.
As the key step of Algorithm~\ref{Alg_Opt} is similar to Step~\ref{algo_MMF_feasibility} of Algorithm~\ref{Alg_Opt_Max_Min_Fair_Problem}, the performance evaluation of Algorithm~1 is omitted here due to space limitation.
We assume channels are i.i.d. Gaussian variables with zero mean and unit variance.
We generate 100 channel realizations and show the performance
in terms of the average worst-case user rate, which is a more meaningful metric than the average worst-case SINR~\cite{chris2014}.
Assume that $N_g,g\!\in\!\mathcal{G}$ are equal.
We choose $\xi_1=10^{-3}$, $\xi_2=10^{-3}$,
${{\mathbf{C}}_{ig}}=1/\mu^{2}{{\mathbf{I}}_{M}}$ and $\sigma^2_{ig}=1,i\!\in\!\mathcal{N}_g,g\!\in\!\mathcal{G}$, $\gamma=M$ Watt in the SP constraints, $\gamma=1$ Watt in the PAP constraints.
We consider three baseline schemes, i.e., the non-robust MM-based method (which treats the estimated channel as the true channel when designing the beamformer)
by extending the method in~\cite{tao2016} and the SDR-based method in~\cite{chen2012} in the case of SP or in~\cite{chris2014} in the case of PAP with the number of Gaussian randomizations $N_\text{rand}$ being $20$ and $100$, respectively.
It is difficult to obtain the worst-case minimum user rate for the non-robust MM-based method which requires solving a non-convex problem with infinitely many constraints $\mathbf{e}_{ig}\in \mathcal{S}_{ig},i\in\mathcal{N}_g,g\in\mathcal{G}$.  Thus for fair comparison, we generate 1000 error realizations for each channel realization and choose the minimum user rate over these error realizations as the worst-case user rate for all schemes, as in~\cite{chris2014}.

\begin{figure}
\centering
\subfigure[$N_g$ under SP constraint. $\mu^2{=}0.25$.]{
\label{fig_rate_vs_Ng_SPC}
\includegraphics[width=1.66 in]{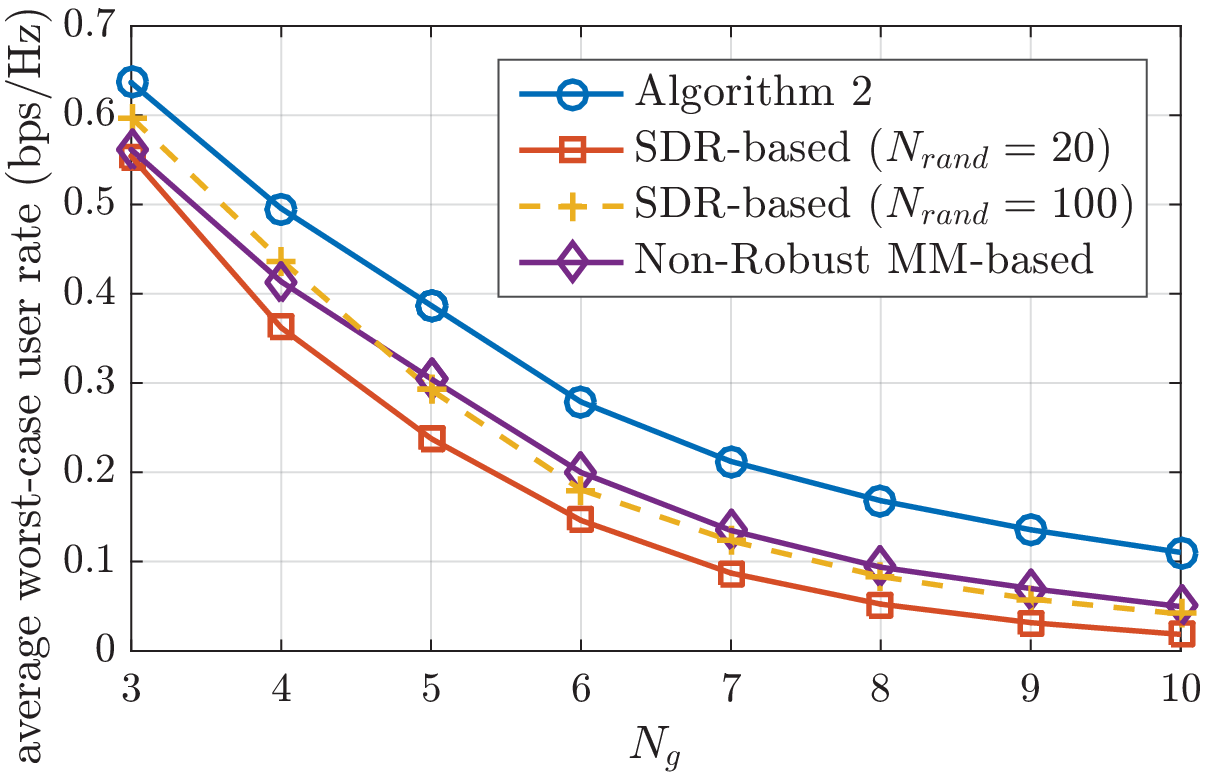}}
\subfigure[$\mu^2$ under PAP constraints. $N_g{=}2$.]{
\label{fig_rate_vsError_PAPCs}
\includegraphics[width=1.68 in]{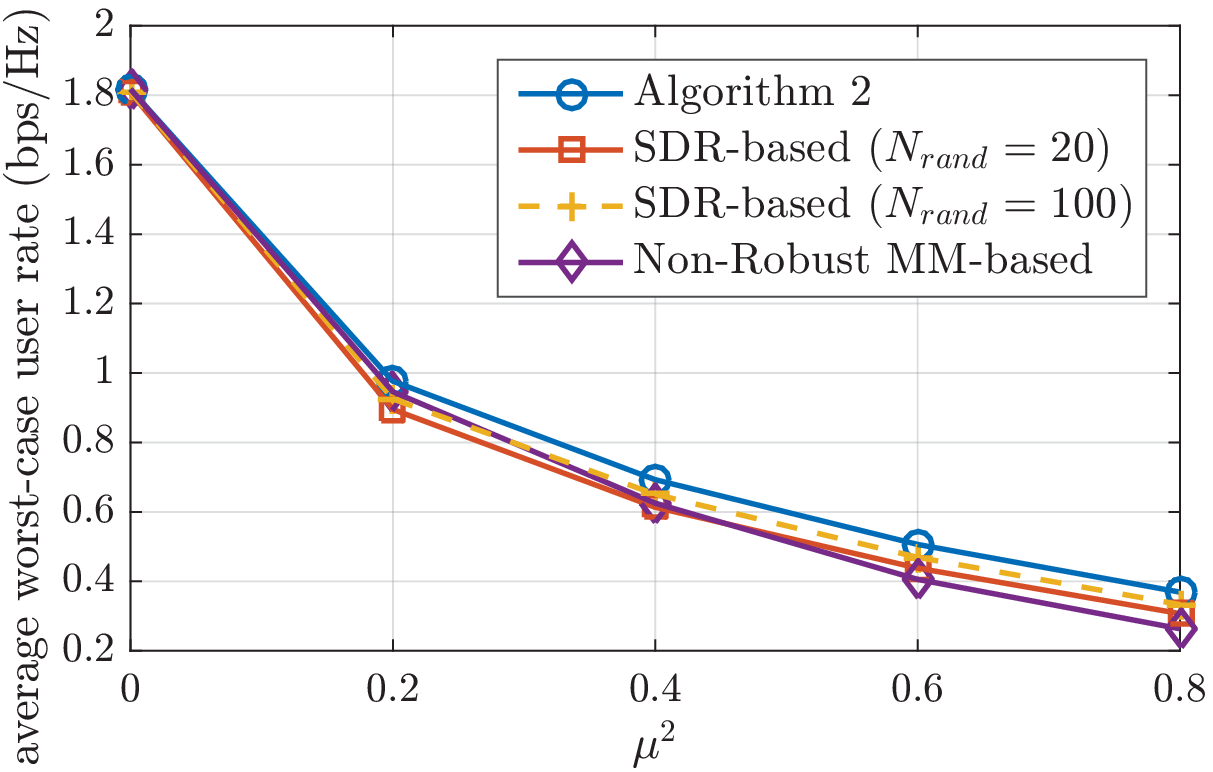}} 
\vspace{-0.4cm}
\caption{Average worst-case user rate versus $N_g$ and $\mu^2$. $M=4$ and $G=2$.}\label{fig_rate_vs_error}
\vspace{-0.1cm}
\end{figure}

\begin{figure}
\centering
\subfigure[SP constraint.]{
\label{fig_complexity_SPC}
\includegraphics[width=1.672 in]{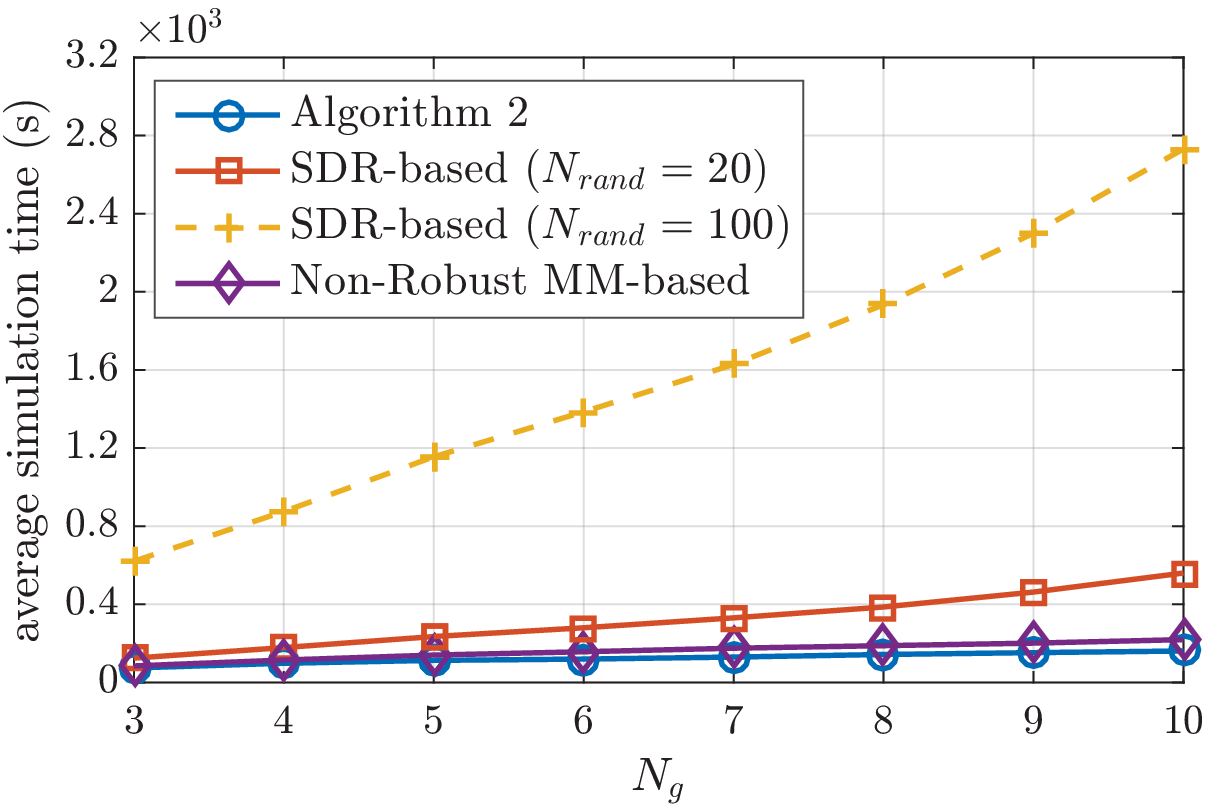}} 
\subfigure[PAP constraints.]{
\label{fig_complexity_PAPCs}
\includegraphics[width=1.67 in]{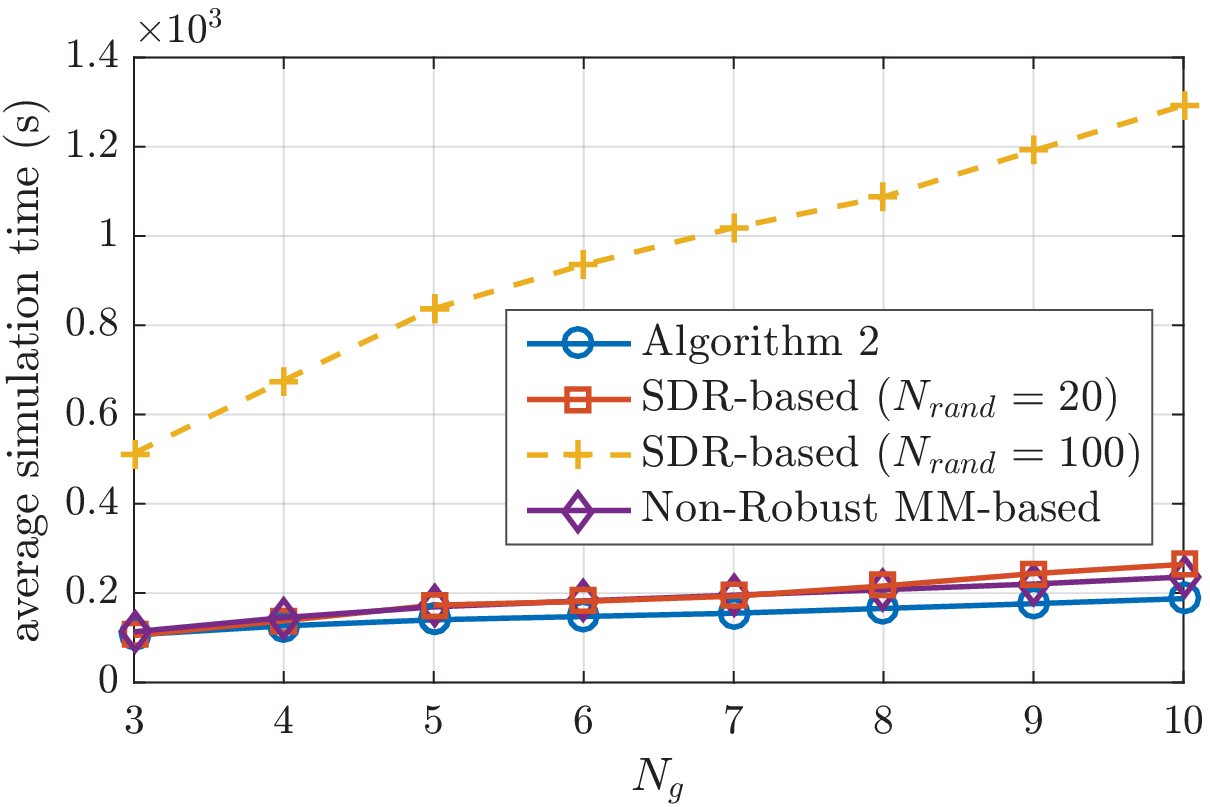}} 
\vspace{-0.4cm}
\caption{Average simulation time versus $N_g$. $M\!=\!4$, $G\!=\!2$ and $\mu^2\!=\!0.25$.}\label{fig_complexity}
\end{figure}

Fig.~\ref{fig_rate_vs_Ng_SPC} and Fig.~\ref{fig_rate_vsError_PAPCs} illustrate the average worst-case user rate versus the number of users per group $N_g$ under the SP constraint and versus the CSI error $\mu^2$ under the PAP constraints, respectively. We can observe that Algorithm~\ref{Alg_Opt_Max_Min_Fair_Problem} outperforms the robust SDR-based method and the non-robust MM-based method.
The gain of Algorithm~\ref{Alg_Opt_Max_Min_Fair_Problem} over the non-robust MM-based method reveals the importance of robust beamforming design at the presence of CSI errors. The gain of Algorithm~\ref{Alg_Opt_Max_Min_Fair_Problem} over the robust SDR-based method demonstrates the value of effective algorithm design for robust optimization.
Fig.~\ref{fig_complexity} illustrates the simulation time (reflecting computational complexity) versus the number of users per group $N_g$. The computational complexity of Algorithm~\ref{Alg_Opt_Max_Min_Fair_Problem} is much lower than that of the
SDR-based method with $N_\text{rand}=100$, and is comparable to those of the SDR-based method with $N_\text{rand}=20$ and the non-robust MM-based method. In addition, in
all of our experiments, the outer loop of Algorithm~\ref{Alg_Opt_Max_Min_Fair_Problem} terminates within 10 iterations.
Therefore, the numerical results demonstrate the advantages of the proposed Algorithm~\ref{Alg_Opt_Max_Min_Fair_Problem} in terms of both the SINR (user rate) and computational complexity.

\section{Conclusion}
In this letter, we considered the robust multigroup multicast beamforming optimizations. The proposed MM-based algorithms can obtain feasible solutions with performance guarantee under certain conditions and low computational complexity.
Numerical results demonstrate the advantages of the proposed ones over existing SDR-based methods.

\ifCLASSOPTIONcaptionsoff
  \newpage
\fi

\bibliographystyle{IEEEtran}
\bibliography{paper}

\end{document}